\documentclass[twocolumn,epsf,psfig]{revtex4-1}
\usepackage{amssymb}
\usepackage{epsfig}
\DeclareMathAlphabet{\pazocal}{OMS}{zplm}{m}{n}            
\usepackage{graphicx}
\usepackage{color}
\usepackage{bm}
\usepackage[normalem]{ulem}     
\usepackage{soul}
\usepackage{amsmath}
\usepackage{braket} 
\usepackage{rotating}
\usepackage{multirow}   
\DeclareSymbolFont{matha}{OML}{txmi}{m}{it}
\DeclareMathSymbol{\varv}{\mathord}{matha}{118}

\begin{document}
\title{Orbital Hall effect as an alternative to valley Hall effect in gapped graphene}         

\author{Sayantika Bhowal} 
\altaffiliation[Current Address: ]
{Materials Theory, ETH Zurich, Wolfgang-Pauli-Strasse 27, 8093 Zurich, Switzerland}
\affiliation{Department of Physics \& Astronomy, University of Missouri, Columbia, MO 65211, USA}

\author{Giovanni Vignale}
\affiliation{Department of Physics \& Astronomy, University of Missouri, Columbia, MO 65211, USA}

\date{\today}

\begin{abstract}
{
Gapped graphene has been proposed to be a good platform to observe the valley Hall effect,  a transport phenomenon involving the flow of electrons that are characterized by different valley indices.  In the present work, we show that this phenomenon is better described as an instance of the orbital Hall effect, where the 
ambiguous ``valley'' indices are replaced by a physical quantity, the orbital magnetic moment, which can be defined uniformly over the entire Brillouin zone. This description  removes the arbitrariness  in the choice of arbitrary cut-off for the valley-restricted integrals in the valley Hall conductivity, as the conductivity in the orbital Hall effect is now defined as the Brillouin zone integral of a new quantity, called the orbital Berry curvature. 
This reformulation in terms of OHE provides the direct explanation to the accumulated  opposite orbital moments at the edges of the sample, observed in previous Kerr rotation measurements.
}

\end{abstract}

\maketitle

\section{Introduction}
The two inequivalent but degenerate valleys in the electronic structure of graphene are used to characterize the electrons by associating them with the ``valley'' index in the same spirit as the usual spin indices are used to designate the electrons. Such valley degrees of freedom in graphene are widely explored \cite{Xiao2007,Schaibley2016,NatNanoLett2018,Song2019} with an eye towards potential use in transport phenomena analogous to the spin transport    
\cite{NatNanoLett2018,Rycerz2007,Vitale2018}. 
One of the notable outcomes of these  studies is the valley Hall effect \cite{Xiao2007},
which is most easily visualized in graphene with broken inversion ($\cal I$) symmetry~\cite{Zhou,Jeon2013}, also known as ``gapped graphene".  The broken $\cal I$ symmetry gives rise to equal and oppositely directed Berry curvatures in the two valleys, giving rise to oppositely directed anomalous velocities in response to an electric field.  As a result, electrons in the two valleys flow in opposite directions, thereby leading to the valley Hall effect (VHE).

\begin{figure}[h] 
\centering
\includegraphics[width=\columnwidth]{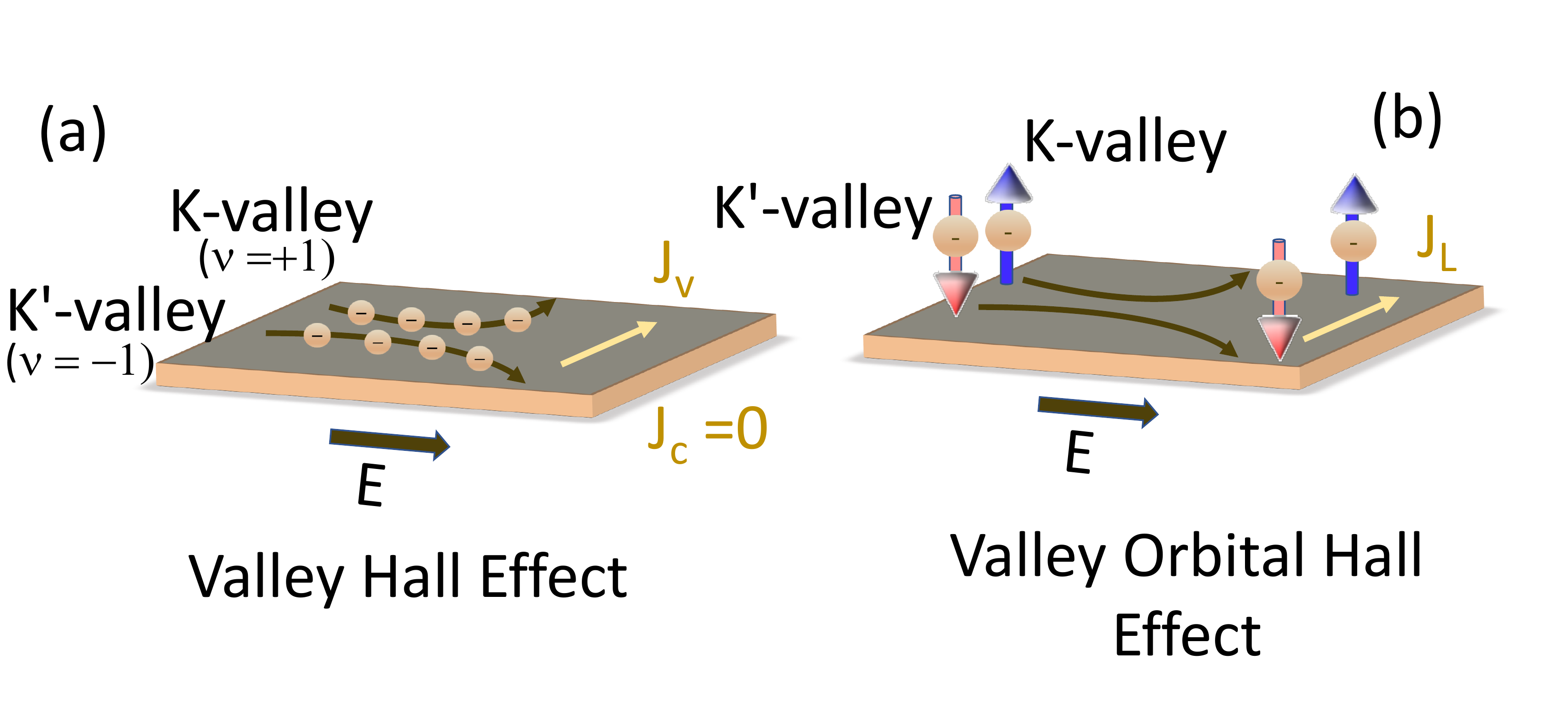}
\caption {Illustration of (a) valley and (b) valley orbital Hall effect, indicating the flow of electrons with opposite valley indices and opposite orbital angular momenta respectively. Notice that due to equal number of electrons in the two valleys, the net transverse charge current ($J_c$) is zero in the intrinsic system. This is in contrast to the  orbital angular momentum current ($J_L$) in the valley orbital Hall effect, which is present even in the intrinsic system.  The big arrows in (b) denote the component of the orbital moment $m^z$, relevant for gapped graphene.  
}
\label{fig0} 
\end{figure}

In the present work, we show that an alternative way to describe the valley degrees of freedom in gapped graphene is to associate the two valleys with the corresponding orbital degrees of freedom of the electrons, i.e., characterizing the 
valley electrons by their orbital angular momenta  rather than the valley indices. 
The two valleys $K, K^\prime$ carry electrons with opposite orbital angular momentum, which also flow in opposite directions due to opposite anomalous velocities at different valleys in the presence of a longitudinal electric field. This leads to 
a transverse orbital Hall current, as schematically illustrated in Fig. \ref{fig0}.  We describe this effect as ``valley orbital Hall effect'' (VOHE).

 This re-description in terms of VOHE is physically more meaningful than the usual  description in terms of VHE, in the sense that the orbital Hall current can be computed uniformly over the entire BZ and the corresponding conductivity is defined as the BZ integral of a new quantity called ``orbital Berry curvature''. This is in contrast to the VHE
 where the conductivity is 
 computed by integrating over a neighborhood of each valley minimum separately, and subtracting, rather than adding, the contributions of the two valleys:
$\sigma^{v, K (K^\prime)}_{xy} = -\frac{e^2}{h}\int_{K (K^\prime)} \frac{d^2k}{2\pi} \sum_n f_{nk} \Omega_{n,z} (\vec k)$ \cite{Souza}.  This definition is not entirely satisfactory, because the valley-restricted integrals $\int_{K (K^\prime)} \frac{d^2k}{2\pi}$ require the demarcation of the two valley domains, which is quite arbitrary.  
 A frequently used prescription  is to consider the $\Gamma-M$ line in the Brillouin zone (BZ) as the boundary between two valleys \cite{Souza, Martiny}.  However, this choice seems to be suggested by practical rather than fundamental considerations.  
 Furthermore, the very sign of the valley Hall current is purely conventional, since we could use $j_K-j_{K'}$ as well as $j_{K'}-j_K$, where $j_{K'}$, $j_K$ are respectively the current densities  corresponding to the individual $K, K'$ contributions to the conductivities $\sigma^{v, K (K^\prime)}_{xy} $. This again casts doubt on the fundamental significance of the ``valley current". 

 On the other hand, the central quantity in the OHE is ``orbital Berry curvature'', which, as we show, is nothing but a product of the orbital moment and the Berry curvature for a gapped graphene system. 
The concept of  ``orbital Berry curvature'' has the advantage that, by definition, it has the same sign at the two valleys, whereas  the ordinary Berry curvature -- the key quantity in the VHE --   has opposite signs at the two valleys. This practically resolves the difficulty of the VHE, i.e., we no longer need to consider an ambiguous ``valley region" in  momentum space.  The magnitude of the computed orbital Hall conductivity does not depend on some arbitrary  momentum cut-off  or valley labelling convention and can be unambiguously compared to the experimental measurements.

In fact, the description in terms of orbital magnetic moments is directly connected with the main experimental technique that is used to detect the VHE, namely Kerr rotation microscopy~\cite{Mak2014,Lee}, in which an out-of-plane polarization with opposite signs at the edges of the sample has been detected in presence of a longitudinal electric current.
 Our redefinition of the effect allows to explain the observed opposite polarizations at the  two edges in terms of the accumulated opposite orbital moments at the edges of the sample due to OHE, emphasizing further the relevance of orbital moments over the abstract ``valley indices".

It is important to notice that the VOHE is only a special case of the more general orbital Hall effect (OHE), which is ubiquitous in solids and does not require broken inversion symmetry \cite{Bernevig,Tanaka,Kotani2008,Kotani2009,Go2018,CysnePRL}. 
In this context, $p$-doped fully hydrogenated {\em centrosymmetric graphene}, known as graphane, is a relevant example in which the OHE, related to the  $2\pi$ Berry flux of the light and heavy hole bands, has already been predicted \cite{Tokatly}.
  The VOHE is specific to 
 non-centrosymmetric systems (like gapped graphene) where each of the two valleys carries an orbital moment due to broken $\cal I$ symmetry, and moments in the two valleys are oppositely directed due to time-reversal symmetry.  Interestingly, as shown in the example of gapped graphene, which we work out in detail below, VOHE does not even require the presence of   spin-orbit coupling.  Thus, we believe that gapped graphene, 
which is usually not considered  a good candidate for spintronics applications due to weak spin-orbit effects \cite{prb2007,prb2006},
could play an important role in the context of  efforts to control the orbital degrees of freedom in two-dimensional systems  by electrical means~\cite{Phong,Zhu2020,prb2020, Canonico,Canonico2020,Xue,TBG2020,OGME2020,Cysne}.  
 The recent experiments \cite{Zheng,Kim2021} on orbital dynamics that utilize  the orbital torque, generated by OHE  \cite{GoPRR,GoPRR2020} in magnetic multilayers, further, raise the hope for future {\em orbitronics} devices.

The rest of the paper is organized as follows. In section \ref{sec1}, we describe the physics of the VOHE using the continuum valley model. We compute the analytical forms for the orbital moment operator, orbital Berry curvature, and the valley orbital Hall conductivity. 
In section \ref{sec2}, the results of the continuum valley model are further verified by using a minimal tight-binding model. The tight-binding model provides the platform to study the VOHE over the entire BZ and the results are compared with those of the continuum model. Our results are discussed and summarized in section \ref{sec3}.

\section{Valley model and VOHE}\label{sec1}

In this section, we illustrate the VOHE for a prototypical massive Dirac Hamiltonian, which is relevant for the valley points $K (4\pi/3\sqrt{3}a, 0), K' (-4\pi/3\sqrt{3}a, 0)$  of the gapped graphene system, e.g., graphene on hexagonal boron nitride. Here $a$ is the distance between C atoms.
 The Hamiltonian near the valley points has the following form
\begin{eqnarray}     \label{Hq}  
{\cal H} (\vec q ) = v_F( \varv \tau_x q_x-\tau_y q_y)+\Delta \tau_z ,
 \end{eqnarray} 
where $\vec q$ represents the crystal momentum near the two valleys, viz.,  $\vec q= \vec k -\vec K$ or $\vec q = \vec k -\vec K'$, $v_F=-3ta/2$ is the Fermi
velocity divided by $\hbar$, which includes the nearest neighbor hopping $t$ between C atoms, $\vec \tau$ are the Pauli matrices in the pseudo spin basis, $2\Delta$ is the energy gap, induced by broken $\cal I$ symmetry, and $\varv$ is the valley index, which can take two values  $\pm 1$ for the $K$, and $K'$ valleys respectively. 

The energy eigenvalues of the valley Hamiltonian (\ref{Hq}) represent gapped Dirac bands with energies  $\varepsilon_{\pm}=\pm \epsilon(q)=\pm \sqrt{\Delta^2+v_F^2q^2}$, where $\pm$ correspond to the conduction and the valence bands respectively. The corresponding wave functions are given by,
   %
  %
\[ \ket{\psi_+}  =\left( \begin{array}{c}
        \varv u_q \\		
        v_qe^{-i\varv\phi} 		
        \end{array} \right), 
  \ket{\psi_-}= \left( \begin{array}{c}
                 v_q\\
                 -\varv u_q e^{-i\varv\phi}
                \end{array}  \right) . \label{psi}
\] 
%
Here $u_q=\sqrt{\frac{1}{2}\big(1+\frac{\Delta}{\epsilon(q)}\big)}$, and $v_q=\sqrt{\frac{1}{2}\big(1-\frac{\Delta}{\epsilon(q)}\big)}$, $\phi = \tan^{-1}(q_y/q_x)$.

The valley Hamiltonian allows for an analytical derivation of the VOHE, which provides useful insight into the physics of the effect.
We start by  computing the matrix elements of the orbital magnetic moment operator for the valley Hamiltonian, which is the key element for the valley orbital Hall conductivity. Before going into the explicit calculation, we want to emphasize that the existence of the non-zero orbital magnetic moment for the Hamiltonian (\ref{Hq}) can be argued even from the simple symmetry transformation relations of the orbital moment $\vec m$, viz., under inversion $\vec m (\vec k) \overset{\cal I}{\rightarrow} \vec m (-\vec k)$, while under time-reversal $\vec m (\vec k) \overset{\cal T}{\rightarrow} -\vec m (-\vec k)$. Therefore, the existence of non-zero $\vec m (\vec k)$ requires either of the two symmetries to be broken. Since the valley Hamiltonian (\ref{Hq}), breaks the $\cal I$ symmetry (because $\tau_z \overset{\cal I}{\rightarrow}-\tau_z$),
 we expect the presence of a non-zero $\vec m (\vec k)$ in the BZ of the system. 

In order to explicitly compute the orbital moment, we construct the corresponding operator, which may be written as
  \begin{eqnarray} \label{om}
 \hat{\vec m} &=& -\frac{e}{4} [ (\hat{ \vec r} \times \hat{ \vec v}) -  (\hat{\vec v} \times \hat{\vec r})] .
 \end{eqnarray}
 Here $\hat{ \vec r} $ is the position operator and $\hat{ \vec v}$ is the velocity operator. 
The gauge-covariant form of the position operator $\hat{ \vec r} $ in the Bloch momentum representation may be computed using the following relation
 \begin{eqnarray}\label{rq}
 (\hat {\vec r})_{qq'} = (i\partial_{\vec q} \delta_{qq'}) \tau_0 + \vec A (\vec q) \delta_{qq'},
 \end{eqnarray}
 where $\vec A (\vec q)$ is the Berry connection, the matrix elements of which can be computed from the eigen states of the Hamiltonian (\ref{Hq}), viz., $A_{nn'} (\vec q) = i \langle \psi_n(\vec q) | \partial_{\vec q} \psi_{n'} (\vec q) \rangle$. By performing this straightforward algebra, we get
  \begin{eqnarray} \label{A}
 \vec A (\vec q)  &=& {\hat q} \Big(\frac{\varv\Delta v_F}{2\epsilon_q^2} \Big) \tau_y + \hat \phi \Big(\frac{v_F}{2\epsilon_q} \Big) \tau_x.
   \end{eqnarray}
 Substituting   $\vec A (\vec q) $ back in Eq. (\ref{rq}), we get the position operator for the Hamiltonian (\ref{Hq}), 
 \begin{eqnarray} \nonumber\label{rH}
 (\hat {\vec r})_{qq'} = 
(i\partial_{\vec q} \delta_{qq'}) \tau_0 + \Big[{\hat q} \Big(\frac{\varv\Delta v_F}{2\epsilon_q^2} \Big)\tau_y + \hat \phi \Big(\frac{v_F}{2\epsilon_q} \Big) \tau_x \Big] \delta_{qq'}. \\
 \end{eqnarray}
All that remains is to construct the velocity operator for the valley model, which is given by
 \begin{eqnarray}\label{vel}
 \hat{\vec v} = \frac{i}{\hbar} [ {\cal H}, \hat {\vec r}] = \frac{1}{\hbar} \Big[ \Big( \frac{\varv\Delta v_F}{\epsilon_q} \Big) \tau_x \hat q -v_F  \tau_y \hat \phi \Big].
 \end{eqnarray}
 Here in obtaining the second equality we have used $ {\cal H} (\vec q)= \epsilon_q  \tau_z$. Using Eqs. (\ref{rH}) and (\ref{vel}), from Eq. (\ref{om}), we get the desired orbital magnetic moment operator for the valley model, 

 \begin{eqnarray} \label{omq}
\hat{\vec m} =\frac{e\varv\Delta v_F^2}{2\hbar\epsilon_q^2}  \tau_0 \delta_{qq'} \hat z = \frac{e}{2\hbar} \frac{\varv v_F^2\Delta}{(\Delta^2 + v_F^2q^2)}  \tau_0 \delta_{qq'} \hat z .
\end{eqnarray}
Note the important result that, for the valley model, the orbital moment matrix is diagonal, i.e., the inter-band elements do not contribute. Also, it is clear that the orbital  moment for the valence and the conduction bands are equal in magnitude, $m^z_v(\vec q) = m^z_c(\vec q) =  \varv \frac{e\hbar}{2m^*} \frac{1}{(1+v_F^2q^2/\Delta^2)}$, where $m^* = \hbar^2 \Delta/v_F^2$ is the effective mass. As expected, due to the presence of time-reversal ($\cal T$) symmetry, the orbital magnetic  moments at different valleys $K,K'$ have opposite signs, resulting in a net zero orbital magnetization in the system. 

In presence of an electric field, these oppositely oriented orbital moments flow in opposite directions, leading to a transverse flow of orbital current. The corresponding valley orbital Hall conductivity for the valence band may be computed as the BZ sum of the orbital Berry curvature $\Omega^{\gamma,\rm orb}_{\alpha \beta} (\vec q)$,
\begin{equation}\label{OHC} 
\sigma^{\gamma,\rm orb}_{\alpha \beta}   =  -\frac{ e } {(2\pi)^2} \int_{\rm BZ}  ~\Omega^{\gamma,\rm orb}_{\alpha \beta} (\vec q)~d^2q,
\end{equation}
where the integration is over the occupied states in the two-dimensional BZ and $-e < 0$ is the electronic charge.
Note the important differences between the orbital Hall conductivity and  the valley Hall conductivity: (i) The integration in Eq.~(\ref{OHC}) is over the entire BZ rather than just around the valley points. (ii) The central quantity in the VOHE is the orbital Berry curvature which has different symmetry properties than the Berry curvature in the valley Hall conductivity, as we discuss now. 

The orbital Berry curvature in Eq.~(\ref{OHC}) may be computed using the Kubo formula,   
\begin{equation}\label{OBC} 
  \Omega^{\gamma,\rm orb}_{\alpha \beta} (\vec q)   =  2\hbar  {\rm Im} \Big( \sum_{n^\prime \neq n} \frac {\langle \psi_{n{\vec  q}} | \mathcal{J}^{\gamma,\rm orb}_\alpha | \psi_{n^\prime{\vec  q}} \rangle  
                        \langle \psi_{n^\prime{\vec  q}} | v_\beta | \psi_{n{\vec  q}} \rangle} 
                        {(\varepsilon_{n^\prime \vec q}-\varepsilon_{n  \vec q})^2 - (i\delta)^2} \Big),
\end{equation}
where the velocity operator $v_\alpha$ in the Berry curvature is replaced by the orbital current operator $\mathcal{J}^{\gamma,\rm orb}_\alpha = \frac{1}{2} \{v_\alpha, L^\gamma \} = \frac{1}{2} (v_\alpha L^\gamma + L^\gamma v_\alpha) $, and $(\alpha, \beta, \gamma) $ are cartesian indices $(x,y,z)$.
The orbital angular momentum operator $L^\gamma = -(\hbar/g_L\mu_B) m^\gamma$, where $g_L=1$ is the Land\'e g-factor for the orbital angular momentum, and $\mu_B =\frac{e\hbar}{2m_e}$ is the Bohr-magneton. Notice that $L^\gamma $ is defined in terms of the orbital magnetic moment and therefore generally differs from the canonical orbital angular momentum operator 
 $(\hat {\vec r} \times \hat {\vec p})^\gamma$, the generator of rotations with quantized eigenvalues. It is also  important to emphasize that the orbital magnetic moment (see Eq. \ref{omq}) in the present gapped graphene model originates from the inter-site circulation current \cite{Xiao,Thonhauser,Shi,Yoda}, the intra-site contribution being zero due to the assumed $s$-like localized orbital basis of our tight-binding model,  discussed later in Eq. \ref{HK}. This is in contrast to the previous studies of OHE on multi-orbital systems \cite{Tanaka,Kotani2008,Kotani2009,Go2018}, where the orbital magnetic moment develops from the intra-site circulation current at each atom.

The parameter $\delta$ in Eq. (\ref{OBC}) represents the finite lifetime broadening that takes into account the impurity scattering effect \cite{Yao2005}. For simplicity, we assume $\delta =0$ here. The lifetime broadening, however, plays an important role in the metallic limit $\Delta \rightarrow 0$, which we discuss later. 

Since for the gapped graphene, only the $z$-component of the orbital magnetic moment is non-zero [see Eq. (\ref{omq})], it is easy to see from Eqs. (\ref{OHC}-\ref{OBC}), that the relevant component of the orbital Hall conductivity  is $ \sigma^{z,\rm orb}_{xy}= -\sigma^{z,\rm orb}_{yx}$. 
 Now, as shown in Eq. (\ref{omq}), the orbital magnetic moment operator is diagonal in the valence ($\psi_-$) and the conduction ($\psi_+$) band representation, i.e., the inter-band matrix elements of the orbital magnetic moment operator $\langle \psi_- | \vec{\hat m} | \psi_+ \rangle = \langle \psi_+| \vec{\hat m} | \psi_-\rangle = 0$, and only the intra-band matrix elements have non-zero values, viz.,  $\langle \psi_- | \vec{\hat m} |\psi_- \rangle = m^z_v (\vec q)$ and $ \langle \psi_+ | \vec{\hat m} | \psi_+ \rangle = m^z_c(\vec q)$, where $m^z_v (\vec q)= m^z_c (\vec q)= \varv \frac{e\hbar}{2m^*} \frac{1}{(1+v_F^2q^2/\Delta^2)}$, as already discussed above. The same is true for the orbital angular momentum as  the orbital angular momentum and the orbital moment differ only by a constant factor.  This essentially simplifies the orbital current operator, the matrix elements of which may be written as
%
\begin{eqnarray}\label{currentOperator} \nonumber
&&(\mathcal{J}^{z,\rm orb}_\alpha)_{nn'} = -\frac{\hbar}{2g_L\mu_B} \langle \psi_{n\vec  q} |  (\hat v_\alpha \hat m^z + \hat m^z \hat  v_\alpha) | \psi_{n^\prime \vec  q} \rangle \\ \nonumber 
 &=& -\frac{\hbar}{2g_L\mu_B}  \big [ \sum_m  \langle \psi_{n\vec  q} | \hat m^z|\psi_{m \vec  q} \rangle  \langle \psi_{m\vec  q} |\hat v_\alpha |\psi_{n' \vec  q} \rangle  \\ \nonumber 
 & & +   \sum_m  \langle \psi_{n\vec  q} |\hat v_\alpha|\psi_{m \vec  q} \rangle  \langle \psi_{m\vec  q} |\hat m^z |\psi_{n' \vec  q} \rangle \big] \\
&=& -\frac{\hbar}{2g_L\mu_B}  \sum_m  [ (\hat m^z)_{nm} (\hat v_\alpha)_{mn'} + (\hat v_\alpha)_{nm} (\hat m^z)_{mn'}] \\
%
  &=& -\frac{\hbar}{2g_L\mu_B} [m^z_n (\vec q)  + m^z_{n'} (\vec q)] \langle \psi_{n\vec  q} |v_\alpha|\psi_{n' \vec  q} \rangle.  \label{eq11}
\end{eqnarray}
%
 Notice that while Eq. (\ref{currentOperator}) is true in general, Eq. (\ref{eq11}) is specific to the Hamiltonian of Eq. (\ref{Hq}), as in obtaining
the last equality, we have used the fact that for this model  $(\hat m^z)_{nm} = \langle \psi_{n\vec  q} | \hat m^z |  \psi_{m\vec  q} \rangle = m^z_n \delta_{nm}$, where $m^z_n =\langle \psi_{n\vec  q} | \hat m^z |  \psi_{n\vec  q} \rangle$ is the orbital moment for the eigenstate $ \psi_{n\vec  q} $ of the Hamiltonian (\ref{Hq}). Here $n = \pm$ correspond to the conduction and valence bands respectively and $\alpha$ represents the Cartesian directions $x,y$. 
In general,  the magnetic moment operator does have off-diagonal elements, which means that Eq.~(\ref{currentOperator}), rather than Eq. (\ref{eq11}), should be used to calculate the OHE.  In the following we focus on the simple model of Eq. (\ref{Hq}).

Using Eqs. (\ref{OBC}) and (\ref{eq11}),  it is  easy to see that the orbital Berry curvature for the valence band of the valley model has the simple analytical form,
\begin{eqnarray}\label{obc} \nonumber
  \Omega^{z,\rm orb}_{xy}(\vec q)  
  &= &  -\frac{2\hbar^2}{g_L\mu_B}  {\rm Im} \Big(  \frac {(m_v^z+m_c^z) [\langle \psi_{-} | v_x | \psi_{+} \rangle  
                        \langle \psi_{+} | v_y | \psi_{-} \rangle]} 
                        {4\epsilon_q^2 } \Big) \\ \nonumber
                        &=& \frac{1}{2g_L\mu_B} [m_v^z(\vec q) +m_c^z (\vec q) ]  \Omega_v^z(\vec q) \\
                    &=& \frac{1}{g_L \mu_B} m_v^z (\vec q) \Omega_v^z(\vec q) ,                       
\end{eqnarray}
%
where $\Omega_v^z (\vec q) = -2 \hbar^2 {\rm Im} \Big(  \frac {\langle \psi_{-} | v_x | \psi_{+} \rangle  
                        \langle \psi_{+} | v_y | \psi_{-} \rangle} 
                        {4\epsilon_q^2 } \Big)=-\frac{\varv v_F^2\Delta}{2\epsilon_q^3}$
is the Berry curvature for the valence band. This is one of the central results of the present work. While physically it describes  the flow of the electrons, carrying the orbital angular momentum, due to the anomalous velocity governed by the Berry curvature, it also provides important insight into the symmetry of the orbital Berry curvature. 
As we can clearly see from this equation, the orbital Berry curvature at the two valleys not only have the same magnitude but also have the same sign as both $m^z_v$ and $\Omega_v^z$ change sign at the two valleys. As a result, the BZ integration of the  
orbital Berry curvature gives a non-zero contribution to the valley orbital Hall conductivity, which is given by
\begin{eqnarray} \label{ohcqsum} 
 \sigma^{z,\rm orb}_{xy}   &=&  \frac{e\Delta^3} {2(2\pi)^2} \Big(\frac{m_e}{m^*}\Big) \int_{\rm BZ}   \frac{v_F^2}{(\Delta^2 + v_F^2q^2)^{5/2}} d^2q.  
\end{eqnarray}
An analytical result is obtained by assuming circular Fermi surfaces  of radius $q_h$ centered at the valley points, where $q_h$ is defined by the Fermi energy $E_ F$, viz.,  $E_F = \pm \sqrt{\Delta^2+v_F^2q_h^2}$. Note that $q_h = 0$ corresponds to the case, when the Fermi energy is at the top of the valence band, i.e., within the energy gap. For any 
 Fermi energy, which lies within the valence band, viz., $E_F \le -\Delta$, the  orbital Hall conductivity is given by 
\begin{eqnarray} \label{ohcq} \nonumber
 \sigma^{z,\rm orb}_{xy}  & \approx &  2e\Delta^3 \Big(\frac{m_e}{m^*}\Big) \frac{1}{(2\pi)^2} \int_{q_h}^{\infty}   \frac{ v_F^2 }{2(\Delta^2 + v_F^2q^2)^{5/2}} d^2 q  \\  \nonumber
 & = &  \frac{e\Delta^3} {6\pi} \Big( \frac{m_e}{m^*}\Big) \frac{1}{(\Delta^2+v_F^2q_h^2)^{3/2}} \\  
 &= &\frac{e} {2\pi} \Big( \frac{m_ev_F^2}{3\hbar^2\Delta}\Big) \frac{\Delta^3}{(\Delta^2+v_F^2q_h^2)^{3/2}} .
\end{eqnarray}

Here, the factor of 2 in the numerator takes care of the contributions for both the valleys, $K, K'$ points, which are equal in magnitude. Note that the conductivity is linear in $v_F^2/ \Delta$, indicating that the magnitude of the effect may be tuned by changing the energy gap of the system. Also, it is clear from Eq. (\ref{ohcq}) that the valley orbital Hall conductivity can occur even without the spin-orbit coupling since there is no spin-orbit coupling in our model. 

Following the same procedure as above and also considering the conduction bands, it is straightforward to see that the result for the orbital Hall conductivity in Eq.~(\ref{ohcq}) is also true if the Fermi energy is inside the conduction band, i.e, $E_F \ge \Delta$. Therefore, Eq.~(\ref{ohcq}) represents the orbital Hall conductivity for the gapped graphene Hamiltonian, Eq. (\ref{Hq}), for any arbitrary Fermi level. It is interesting to note that the orbital Hall conductivity has a particle-hole symmetry, as evident from the $q_h^2$ dependence of $\sigma^{z,\rm orb}_{xy} $ in Eq. (\ref{ohcq}). The conductivity is maximum when the Fermi level is within the energy gap ($q_h=0$), where it attains the value $ \sigma^{z,\rm orb}_{xy}(0)=\frac{e} {6\pi} \Big( \frac{m_e}{m^*}\Big)$, and, then, falls off as $E_F$ shifts on either side of the energy spectrum.  Notice that, unlike the valley Hall conductivity,  which  is quantized to the value $e^2/h$, independent of $\Delta$,  when the Fermi energy is within the energy gap~\cite{Souza}, the orbital Hall conductivity does depend on the energy gap $\Delta$ through the effective mass, which enters  the expression of the magnetic moment.

  \section{Two sub-lattice tight-binding model}\label{sec2}
 The analytical results for the valley orbital Hall effect, discussed above, provide several important insights into the effect. As shown above, in contrast to the valley Hall effect where the two valley contributions are equal and opposite, the two valleys contribute equally to the VOHE, both in magnitude and sign, which allows us to define the valley orbital Hall current  over the entire BZ of the system and not just in the vicinity of the two valleys. To demonstrate this crucial distinction of the VOHE from the valley Hall effect, it is instructive to work out the VOHE for a lattice model, which gives access to the wave functions over the entire BZ and not just around the valley points as in Eq.~(\ref{Hq}).

We start by constructing the tight-binding model for a honeycomb lattice, defined by the lattice vectors $\vec a_1 = (a/2)(-\sqrt 3 \hat i + 3 \hat j)$ and $\vec a_2 = (a/2)(\sqrt 3 \hat i + 3 \hat j)$, with two atoms A and B at $(0,0)$ and $(0,-a)$ respectively.  
For simplicity, we have considered only hopping $t$ between the nearest neighbor (NN) atoms, connected by the three vectors $\vec \delta_1 = (a/2)(-\sqrt 3 \hat i + \hat j)$, $\vec \delta_2 = (a/2)(\sqrt 3 \hat i + \hat j)$, and $\vec \delta_3 = - a \hat j$. The onsite energies of A and B atoms are considered to be different, viz. $\pm \Delta$, which breaks $\cal I$ symmetry and gives us a gapped spectrum. The resulting tight-binding Hamiltonian is
\begin{eqnarray} 
 {\cal H} (\vec k)= \sum_{\vec k, \alpha, \beta} \hat c^\dagger_{\vec k \alpha} H_{\alpha, \beta}(\vec k) \hat c^\dagger_{\vec k \beta},
\end{eqnarray}
where \begin{eqnarray}    \label{HK}  
H_{\alpha, \beta} (\vec k )& =& 
\left[
{\begin{array}{*{20}c}
    \Delta & f(\vec k)  \\
     f^*(k) &  -\Delta\\
   \end{array} }  \right],
   \end{eqnarray} 
with $f(\vec k) = t [1 + 2\cos(\sqrt{3}k_xa/2) \exp (i3k_ya/2)]$.

\begin{figure}[t] 
\centering
\includegraphics[width=\columnwidth]{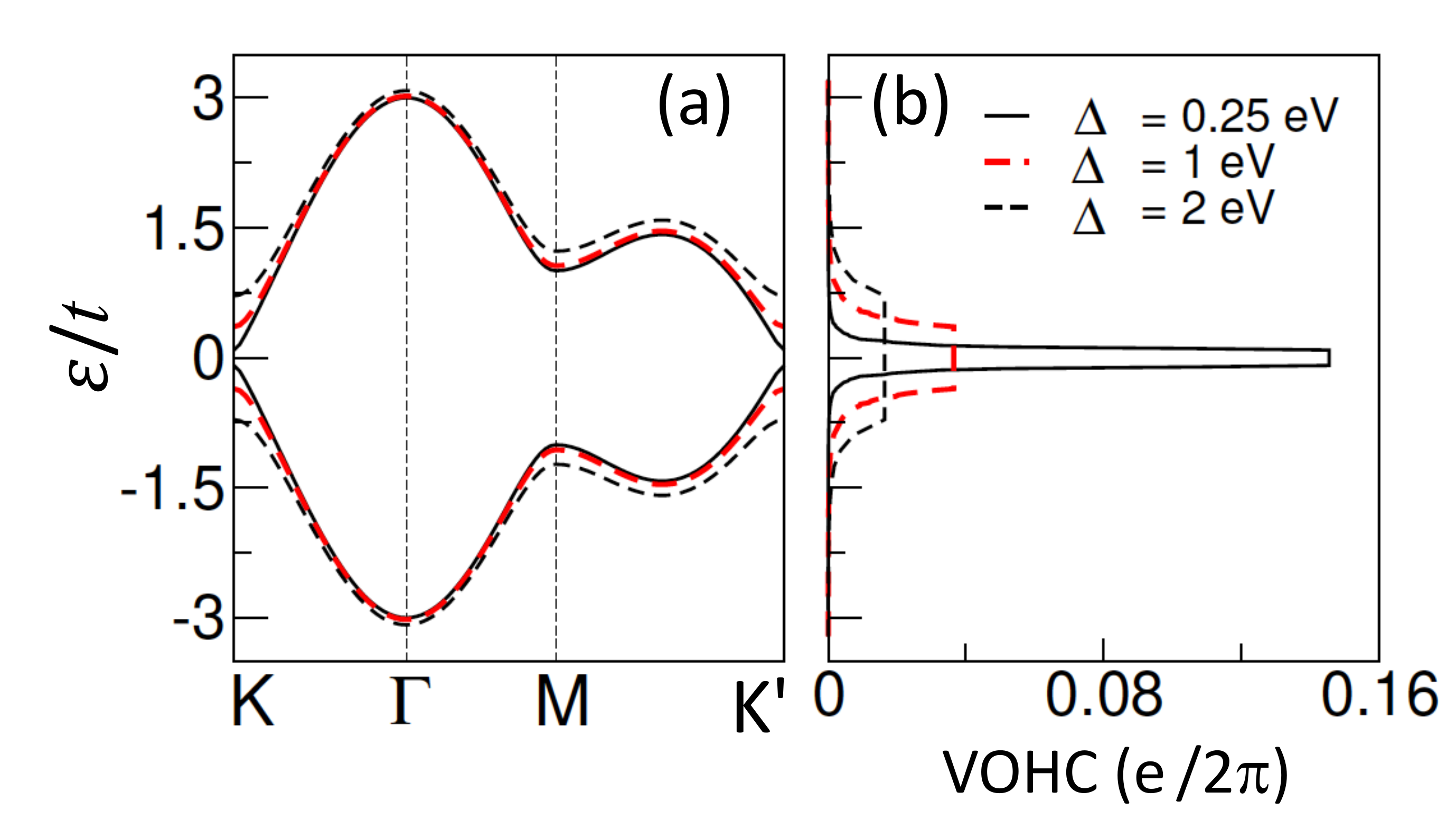}
\caption {(a) The band structure of gapped graphene [see Eq. (\ref{EHK})] for different values of the parameter $\Delta$, that determines the energy gap. (b) The corresponding valley orbital Hall conductivity (VOHC), shown as a function of the Fermi energy. The magnitude of the VOHC is maximum when the Fermi energy lies within the gap. The dependence of the VOHC on the parameter $\Delta$ is also apparent from this plot. The typical values of the chosen parameters are: $t = 2.8$ eV, and $a = 1.42$ \AA. 
}
\label{fig2} 
\end{figure}

The energy eigenvalues of the Hamiltonian (\ref{HK}) are 
\begin{equation}\label{EHK}
\varepsilon_\pm(\vec k) = \pm \sqrt{\Delta^2 + |f(\vec k)|^2},
\end{equation}
where $\pm$ correspond to the conduction ($+$) and the valence ($-$) bands of gapped graphene, as shown in Fig. \ref{fig2} (a). Notice that expanding the Hamiltonian $H(\vec k)$ around the valley points $K (4\pi/3\sqrt{3}a, 0)$ and $K' (-4\pi/3\sqrt{3}a, 0)$, and defining $v_F=-3ta/2$ we can immediately recover the valley model, Eq. (\ref{Hq}), discussed
in the previous section.

\begin{figure}[t] 
\centering
\includegraphics[width=\columnwidth]{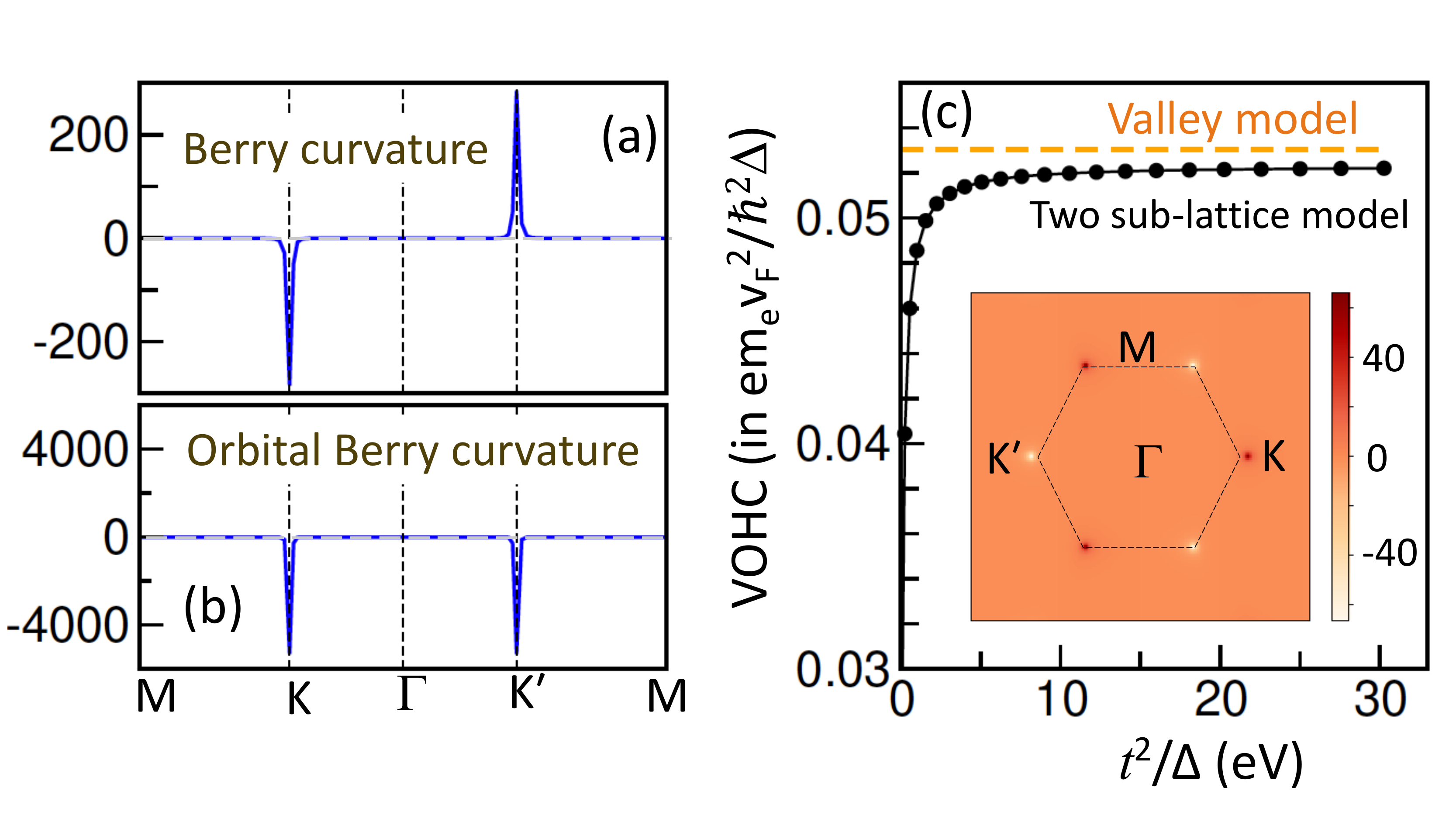}
\caption {(a) The Berry curvature, Eq. (\ref{bck}), and (b) the orbital Berry curvature, Eq. (\ref{obck}),  both in units of \AA$^2$, for the valence band of the Hamiltonian (\ref{HK}) along a high symmetry $k$-path. While the Berry curvature, has opposite signs at $K,K'$ valleys, the orbital Berry curvature at different valleys have the same sign. (c) The variation of the valley orbital Hall conductivity (VOHC) in units of $e m_e v_F^2 (\hbar^2\Delta)^{-1}$ as a function of $t^2/\Delta$, where $v_F \propto t$. The dashed line corresponds to the analytical result, Eq. (\ref{ohcq}), at $q_h=0$ for the valley model, indicating agreement between the two models in the large $t^2/\Delta$ regime, where VOHC varies linearly with $t^2/\Delta$. The inset shows the  distribution of the orbital moment (in units of $e/\hbar$) in the BZ. 
The parameters are the same as stated in the caption of Fig. \ref{fig2}, except in (c), where the hopping $t$ is varied keeping $\Delta$ fixed at 0.25 eV to get the variation of $t^2/\Delta$. For these typical chosen parameters, Berry curvature, orbital moment, and  orbital Berry curvature are localized around the valley points. }
\label{fig3} 
\end{figure}

We now proceed to compute the valley orbital Hall effect for the two sub-lattice model (\ref{HK}).
 Similar to the continuum model, we start by calculating the matrix elements of the orbital magnetic moment operator, which is given by
\begin{eqnarray}\nonumber \label{mnn'}
(\hat {\vec m})_{nn'} &=& \frac{e}{2\hbar} {\rm Im} [ \langle {\partial_{\vec k} u_n(\vec k)}| \times H(\vec k)  \ket{\partial_{\vec k} u_{n'}(\vec k)}] \\
&& + \frac{e}{4\hbar} [\varepsilon_n (\vec k)+\varepsilon_{n'}(\vec k)]  \hat {\vec{\Omega}}_{nn'} .
\end{eqnarray}
Here, the matrix element of the Berry curvature $ \hat {\vec{\Omega}}_{nn'}  =- {\rm Im} [\langle \partial_{\vec k} u_{n} ({\vec k})| \times | \partial_{\vec k} u_{n'} ({\vec k}) \rangle ] $. Using the following identity for the $\vec k$ derivative of the Bloch bands,
\begin{eqnarray}
\ket {\partial_{\vec k} u_n(\vec k)} =\sum_{n' \ne n} \frac{\langle u_{n'} ({\vec k})|\partial_{\vec k} H(\vec k)| u_n ({\vec k})\rangle }{(\varepsilon_{n'}-\varepsilon_n)} \ket{u_{n'} ({\vec k})}
\end{eqnarray}
and recalling that the eigenstates $\ket {u_{\pm} ({\vec k})}$ of the Hamiltonian $H(\vec k)$ are orthogonal to each other, i.e., $\langle u_+ ({\vec k})|  u_- ({\vec k}) \rangle =0$, after some  straight forward algebra, we can see from Eq. (\ref{mnn'}) that the off-diagonal elements ($n \ne n'$) of the orbital moment operator vanishes for the Hamiltonian (\ref{HK}). The diagonal elements are, however, non-zero, and for $n=n'$, we recover the well known formula \cite{Xiao,Shi} for the 
intrinsic orbital moment $\vec m_n$ for the $n^{th}$ band, as given below. 
\begin{eqnarray}\nonumber
 \vec m_n(\vec k) = \frac{e}{2\hbar} {\rm Im} \bra {\partial_{\vec k} u_n(\vec k)} \times [H(\vec k) -\varepsilon_n (\vec k)] \ket{\partial_{\vec k} u_n(\vec k)} . \\
\end{eqnarray}
Explicit calculation shows that the orbital magnetic moment has only non-zero component along $z$-direction, $m^z_n$ which again has the same magnitude for both the valence and the conduction bands of the Hamiltonian (\ref{HK}), viz., 
\begin{eqnarray} \nonumber \label{mz}
 m^z_v(\vec k) = m^z_c(\vec k) = -\frac{e}{4\hbar} \frac{3\sqrt{3}a^2t^2\Delta}{(\Delta^2 + |f(k)|^2)} \sin (\sqrt{3}k_xa). \\
\end{eqnarray}
The momentum space distribution of  $m^z_v$ is depicted in the inset of Fig. \ref{fig3} (c), showing that the magnitude of the orbital moment is maximum at the valley points and it has opposite directions at the two valleys, in agreement with the valley model. 
Using the orbital magnetic moment operator $m^z$, and the velocity operators for the Hamiltonian (\ref{HK}),
\begin{eqnarray}  \nonumber  \label{vx}  
v_x & =& \hbar^{-1}
\left[
{\begin{array}{*{20}c}
    0 & \partial_{k_x} f(\vec k)  \\
     \partial_{k_x} f^\star(\vec k) &  0\\
   \end{array} }  \right],\\
v_y & =&  \hbar^{-1}
\left[
{\begin{array}{*{20}c}
    0 & \partial_{k_y} f(\vec k)  \\
     \partial_{k_y} f^\star(\vec k) &  0\\
   \end{array} }  \right] , \label{vy}
   %
   \end{eqnarray} 
where
$
 \partial_{k_x} f(\vec k) = -\sqrt{3} ta \sin (\sqrt{3}k_xa/2) \exp (i 3k_ya/2), $ and 
 $\partial_{k_y} f(\vec k) =  i3 ta \cos (\sqrt{3}k_xa/2) \exp (i 3k_ya/2)$,
we can compute the orbital Berry curvature using Eq. (\ref{OBC}). 
The result for the valence band is
\begin{equation}\label{obck}
 \Omega^{z,\rm orb}_{xy} ({\vec k})  =  -\frac{2m_e} {g_L\hbar^2}  \frac{(3\sqrt{3}a^2t^2\Delta)^2}{16} \frac{\sin^2 (\sqrt{3}k_xa)}{(\Delta^2 + |f(\vec k)|^2)^{5/2}}.
\end{equation}
Note that the orbital Berry curvature is an even function of $k_x$, in contrast to the 
 Berry curvature, which is an odd function. The explicit analytical expression for the Berry curvature for the valence band of the tight-binding model (\ref{HK}) is given by,
\begin{equation}\label{bck}
 \Omega_v^{z} ({\vec k})  =   \frac{3\sqrt{3}a^2t^2\Delta \sin (\sqrt{3}k_xa)}{4(\Delta^2 + |f(\vec k)|^2)^{3/2}}.
\end{equation}
 Combining Eqs. (\ref{mz}), (\ref{obck}), and (\ref{bck}), we can again see that the orbital Berry curvature is nothing but the product of the orbital moment and the Berry curvature,
\begin{equation}\label{obckn}
  \Omega^{z,\rm orb}_{xy} ({\vec k})  = \frac{1}{g_L \mu_B} m^z_v(\vec k) \Omega_v^{z} ({\vec k}).  
\end{equation}
%

Both the Berry curvature and the orbital Berry curvature are shown 
in Figs. \ref{fig3} (a) and (b) respectively along a high symmetry $k$-path. As expected from the analytical results, the orbital Berry curvature has the same sign at the two valleys in contrast to the Berry curvature, which changes sign at the two valleys. The orbital Berry curvature being an even function in $\vec k$, the sum of the orbital Berry curvature gives a non-zero value and the conductivity  reads as
\begin{eqnarray}\label{ohck} \nonumber
\sigma^{z,\rm orb}_{xy}   &=&  \frac{2m_ee } {g_L\hbar^2 (2\pi)^2} \int_{\rm BZ}  \frac{(3\sqrt{3}a^2t^2\Delta)^2}{16} \frac{\sin^2 (\sqrt{3}k_xa)}{(\Delta^2 + |f(\vec k)|^2)^{5/2}} d^2k, \\
\end{eqnarray}
where the integration is over the occupied BZ. The numerical value of the integration is shown in Fig. \ref{fig2} (b) as well as in Fig. \ref{fig3} (c). Fig. \ref{fig2} (b) depicts the variation of the valley orbital Hall conductivity as a function of the Fermi energy ($E_F$). As we can see from this figure, the valley orbital Hall conductivity is maximum when $E_F$ lies within the gap, while the conductivity falls off as $E_F$ moves into the valence or the conduction band, in agreement with the continuum model result Eq. (\ref{ohcq}). Note that the magnitude of the effect decreases with $\Delta$, which is also expected from the $1/\Delta$ dependence of the orbital Hall conductivity [see Eq. (\ref{ohcq})].

The variation of the valley orbital Hall conductivity in units of $em_ev_F^2 (\hbar^2\Delta)^{-1}$, for the case when the Fermi energy is within the energy gap,  is shown in Fig. \ref{fig3} (c) as a function of $t^2/\Delta$. As seen from Fig. \ref{fig3} (c), the scaled valley orbital Hall conductivity (as it is in units of $e m_e v_F^2 (\hbar^2\Delta)^{-1}$, where $v_F \propto t$) increases for small $t^2/\Delta$ and, then, quickly saturates to a constant value, indicating that in the large $t^2/\Delta$ regime, valley orbital Hall conductivity is linear in $t^2/\Delta$. Similar linear $t^2/\Delta$ dependence can also be seen in Eq. (\ref{ohcq}) for the valley model with $q_h =0$, as well as in Fig. \ref{fig3} (c), showing that the scaled valley orbital Hall conductivity ($\sigma^{z,orb}_{xy}/e m_e v_F^2 (\hbar^2\Delta)^{-1}$) is a constant $1/6\pi \approx 0.053$ 
for all values of $t^2/\Delta$.

We note that for large values of $t^2/\Delta$, the orbital Hall conductivity, computed from the two sub-lattice model, agrees well with the valley model. This can be understood from the fact that in the small $\Delta$ regime, the orbital Berry curvature is  concentrated  around the valley points, and, therefore, in this regime the lattice model essentially reduces to the valley model.

We now analyze the contributions of  different parts of the occupied BZ to  the total orbital Hall conductivity. It is easy to see from Fig. \ref{fig3} (b) that the orbital Berry curvature has a significant magnitude only around the valley points, while it is almost zero everywhere else,  hence we can refer to the effect as ``valley orbital Hall effect". The strong valley contributions to the orbital Hall conductivity may be traced back to the intrinsic orbital magnetic moment $m^z$ in the $k$-space, which, together with the Berry curvature, constitutes the orbital Berry curvature at each $k$ point, as in Eqs. (\ref{obc}) and (\ref{obckn}). While the individual sums of $m^z$ or the Berry curvature over the entire occupied BZ vanish, it is this new quantity, the orbital Berry curvature, which has a non-zero value over the occupied BZ. 

 It is also important to appreciate that the simple analytical form in Eqs. (\ref{obc}) and (\ref{obckn}),  expressing the orbital Berry curvature as the product of orbital moment and  Berry curvature, is strictly a characteristic of the gapped graphene model and may not be the case in general. In other words, the orbital current operator may not have, in general,  the simple form as in Eq. (\ref{eq11}), where only the intra-band matrix elements of the orbital magnetic moment operator contribute. In general, the inter-band matrix elements may also appear and the orbital Berry curvature should be explicitly computed from Eq. (\ref{OBC}) using the general form of the orbital current operator, Eq.  ({\ref{currentOperator}).
 In fact, for centro-symmetric systems where the intra-band matrix elements of the orbital angular momentum operator vanishes, it is the inter-band matrix elements of orbital angular momentum operator that give rise to the non-zero orbital angular momentum current, resulting in an OHE even in presence of inversion symmetry \cite{Go2018,Sahu2021}. Conversely,  for a non-centrosymmetric system \cite{prb2020,Canonico},  the orbital Hall conductivity may not be so tightly connected to the valley points as it is in the gapped graphene. This is precisely because, in general, all the momentum points in the BZ contribute to the OHE, in contrast to the VHE that only considers the contributions from the valley points.   Notice that the intra- and inter-band contributions to the orbital moment is different from the previously discussed intra- and inter-site orbital magnetic moment.

We now consider the $\Delta \rightarrow 0$ limit, i.e., when the Dirac bands touch at the valley points and the Fermi energy lies exactly at the point of contact between the bands. In this limit, the orbital moment vanishes everywhere in the BZ, except at the valley points where the orbital moment as well as the orbital Berry curvature diverges. A similar divergence is also observed in the diverging 
 orbital diamagnetism of graphene. However, 
in the metallic limit, disorder may remove the divergence by shifting the Fermi energy away from the crossing point or by lifting altogether the degeneracy at the band crossing point. In spite of these difficulties, recent experiments were able to observe a diverging orbital diamagnetism in clean graphene monolayers~\cite{arxiv2020}. This gives us ground to hope that a similar divergence  may be observed in experimental measurements of the VOHE as well.  Such a diverging feature is absent in the valley Hall conductivity in the $\Delta \rightarrow 0$ limit.

 \section{Summary and discussion}\label{sec3}

 To summarize, we have suggested that a more physical description of the valley Hall effect in systems like ``gapped graphene" can be given in terms of the orbital Hall effect. Thus, instead of a fictitious ``valley current", which has no physical operator associated with it, and requires the introduction of arbitrary  demarcations in momentum space, we talk of a current of orbital magnetic moment, which has a physical operator associated with it, and can be computed  uniformly and unambiguously over the entire momentum space.   
To be sure, the OHE is a more general concept than the VHE, since a current of orbital magnetic moment can exist even in centro-symmetric systems where it is not associated with valley degrees of freedom.  In the broken inversion symmetry systems considered here the OHE subsumes the VHE: it is to emphasize this tight connection that we have chosen to call it {\it valley}-orbital Hall effect.  

 Nevertheless, the view proposed here has also implication for the understanding of the OHE in certain inversion-symmetric systems, for example, the bilayer of transition-metal dichalcogenide with ``hidden" broken inversion symmetry of the individual layers, studied recently by Cysne {\it et al.} in Ref.~\cite{CysnePRL}. This system has inversion symmetry and therefore does not support a VHE.  The orbital Hall conductivity of the bilayer, when the Fermi level is in the band gap, turns out to be essentially twice the value of the single layer.  But the single layer has broken inversion symmetry and might therefore be suspected of supporting a VHE, which would  contribute to the magnetization current  in addition to the OHE~\cite{CysnePRL}.  In the present view, however, the VHE of the single layer does not produce any additional orbital magnetization current. 
It is the same orbital magnetization current that is generated by the OHE, which is, in fact, the physical effect. Then,  combining  the two layers simply doubles the OHE (as long as the Fermi level lies within the energy gap, suppressing the effect of interlayer hopping).
 
 On a mathematical level, two different  quantities underlie the VHE and the VOHE, viz., the Berry curvature for the VHE and the ``orbital Berry curvature" for the VOHE. We have shown that for both the continuum model as well as the two sub-lattice model of gapped graphene, the orbital Berry curvature acquires a simple form, i.e., it is defined by the product of the  magnetic moment  and the Berry curvature.
 
While the Berry curvature has opposite signs at different valleys, the orbital Berry curvature has the same sign, simply because both the magnetic moment and the Berry curvature change sign in going from one valley to the other.  An interesting consequence of this is the possibility of getting a finite value by integrating the orbital Berry curvature over the entire BZ without any arbitrary assumptions,  whereas the integral of the Berry curvature would vanish.

Our calculation shows that the valley orbital Hall conductivity varies as $t^2/\Delta$, indicating the possibility to tune the effect by changing the energy gap, which may be achieved by growing graphene on different substrates. Furthermore, the Fermi energy can also control the magnitude of the effect. A maximum conductivity may be achieved if the Fermi energy falls within the energy gap. Our work emphasizes the important role of broken $\cal I$ symmetry in the VOHE in gapped graphene. 

Experimentally, VOHE can be detected by measuring the accumulations of orbital magnetic moment at the edges of the sample via magneto-optical Kerr rotation \cite{Kato}. Angle-resolved photo-emission measurements \cite{Park,Beaulieu} can also be used to probe the valley orbital moments.

\section{Acknowledgement}
The authors thank Tarik Cysne and Tatiana Rappoport for useful discussions.
S. B. thanks the U. S. Department of Energy, Office of Basic Energy Sciences, Division of Materials  Sciences and Engineering for financial support under Grant No. DE-FG02-00ER45818.


\end{document}